\documentstyle[12pt]{article}
\newcommand{\Sm}[1]{{\rm S_-^{{#1}}}}
\newcommand{\Sp}[1]{{\rm S_+^{{#1}}}}
\newcommand{\Sl}{{\rm S}}

\newcommand{\Np}{{\cal N}}

\newcommand{\vct}[1]{|{#1}\rangle}

\newcommand{\vac}{|0\rangle}

\newcommand{\bin}[2]{C^{{#1}}_{{#2}}}

\newcommand{\psin}[3]{\psi_{{#1}}({#2}_{1},\ldots,{#2}_{#3})}

\newcommand{\psiN}[5]{\psi_{{#1}}^{l}({#2}_{1},\ldots,{#2}_{#3},{#4}_{1},
\ldots,{#4}_{#5})}

\newcommand{\eps}{\epsilon(l)}
\newcommand{\hd}{\delta{\rm H}}
\newcommand{\al}{\alpha}
\newcommand{\bt}{\beta}
\newcommand{\PSS}[2]{\psi_{S}^{l}({#1},{#2})}
\newcommand{\PsiN}[2]{\psi^{l}({#1},{#2})}
\newcommand{\PhiN}[2]{\phi^{l}({#1},{#2})}
\newcommand{\Psin}[2]{\psi_{{#1}}({#2})}
\newcommand{\Phin}[2]{\phi_{{#1}}({#2})}

\begin{document}
\begin{titlepage}
\hfill\parbox{35mm}{{\sc SPbU--IP--96--8}\par
                              hep-th/9604173 \par
                              April, 1996}
\vskip 3cm
\begin{center}
{\bf \Large
Generic scaling relation
in the scalar
$ \phi^{4} $ model.}
\vskip 1cm
{\bf \large S.~\'{ E}.~Derkachov ${}^{a}$
\footnote{E--mail: derk@tu.spb.ru}
and
 A.~N.~Manashov ${}^{b}$
\footnote[2]{E--mail: manashov@phim.niif.spb.su}
}\\
\vskip 1cm
$ {}^{a} $ {\it Department of Mathematics,
St.-Petersburg Technology Institute, St.-Petersburg, Russia} \\
$ {}^{b} $ { \it Department of Theoretical Physics,  Sankt-Petersburg State
University, St.-Petersburg, Russia }
\vskip 1cm
{\large \today}
\\
\vskip 2cm
{\bf \large Abstract}
\end{center}
\vskip 0.5cm
\begin{center}
\begin{minipage}{14cm}
The results of  analysis of the one--loop spectrum of
anomalous dimensions of composite operators in the scalar
$ \phi^{4} $ model are presented.
We give the rigorous constructive proof
of the hypothesis on the hierarchical
structure of the spectrum of anomalous dimensions~---~the naive sum
of any two anomalous dimensions generates a limit point in
the spectrum.
 Arguments in favor of the
nonperturbative
character of this result and the possible ways of a generalization
to  other field theories are briefly discussed.
\end{minipage}
\end{center}
\end{titlepage}
\newpage

\renewcommand{\theequation}{\thesection.\arabic{equation}}
\setcounter{equation}{0}
\section{Introduction}
\setcounter{page}{1}

\label{intr}

The great interest to the renormalization of  composite
operators is mainly motivated by the successful application of
the operator product expansion (OPE)  method \cite{OPE} for
the description of the processes of  deep
inelastic scattering in QCD. It is
well known now that the behavior of the structure functions at
large $ Q^{2} $ is closely related to the spectrum of
anomalous dimensions of  composite operators~\cite{Hard}.
The mixing matrix  describing the renormalization
of the operators in QCD is well known for any set of the
latters~\cite{Lip}. Unfortunately,
the size of the mixing matrix for the  operators with
given spin
$ j $  and twist greater then 2
increases very fast with the growth of
$ j $, so  the problem of the calculation of  anomalous
dimensions seems to admit only numerical solution for not too large
values of
$ j $.

The first attempt to analyze the spectrum of  anomalous
dimensions for the
particular class of twist--3 operators  in the limit
$j\to\infty $
has been undertaken in the
paper~\cite{Braun}. In this work the analytic solution for the
minimal anomalous dimension has been obtained. The exact solution
as well as the results of the numerical study of the spectrum
allow to suggest that the asymptotic behavior of  anomalous
dimensions of twist--3 operators in some sense is determined  by
the spectrum of those with twist--2.

It should be stressed that the troubles connected with the
calculation of the spectrum of  anomalous dimensions are not
the peculiarity of QCD only. One encounters  the same
problems and in more simple theories.
In the recent
papers~\cite{WP,WK,DM,Kh} the analysis of the spectrum
of  anomalous dimensions in
$ O(N) $~--~vector model in
$ 4-\epsilon $ dimensions has been carried out.
Due to relative simplicity of this model it is appeared
possible to obtain the exact solution of the eigenvalue problem
for some classes of composite operators
(see refs.~\cite{WP,DM}). The consideration of these
solutions together with the results of the numerical analysis
fulfilled in~\cite{WK} for the wide class of operators leads
to the same conclusion that  the asymptotic of  anomalous
dimensions of the operators with given twist in the $ j\to \infty
$ limit is determined by the dimensions of the
operators with a smaller twist~\cite{DM,Kh}.  It was supposed
in~\cite{Kh}
that spectrum of  anomalous dimensions has a~"hierarchical"
structure; this means that the sum of any two points of the
spectrum is the limit point of the latter.

In the present paper we investigate the large spin asymptotic of
 one--loop anomalous dimensions of the  spatially traceless
and symmetric composite operators in the scalar
$ \phi^{4} $ in
\mbox{$ 4-\epsilon $} dimensions. The approach used here is
not specific only for this model and admits the straightforward
generalization for other theories.

Before proceeding with calculations we would like to discuss the
main troubles which arise in the course of the analysis of the
spectrum of anomalous dimensions of  large  spin operators.
It is easy to understand that the source of all difficulties
is the mixing problem.
Indeed, a long time ago C.~Callan and D.~Gross proved a very
strong statement concerning  anomalous dimensions of the
twist-2 operators \cite{CG} for which the mixing problem is
absent.  They obtained that in all orders of the perturbation
theory the anomalous dimension $ \lambda_{l} $ of the operator
$ \phi\partial_{\mu_1}\ldots\partial_{\mu_l}\phi $
tends to
$ 2\lambda_{\phi} $ at
$ l\to\infty $ ($ \lambda_{\phi} $ is the
anomalous dimension of the field $ \phi $).

Let us see what prevents  the generalization of this result,
even on the one loop level,   for the case of the operators
of higher twists.
Although to calculate the mixing
matrix is not very difficult, the extraction of the  information
about  eigenvalues of the latter needs a lot of work.  Really,
if one has not any idea about structure of eigenvectors the
only way to obtain  eigenvalues is to solve a characteristic
equation. But it is almost a hopeless task. However, let us
imagine that one has a guess on the form of an eigenfunction; then
there are no problem with the evaluating the corresponding
eigenvalue.  (Note, that the exact solutions in
\cite{Braun,WP,DM} were obtained precisely in this manner.) Thus
the more promising strategy is to guess the approximate structure
of  eigenfunctions in the  "asymptotic" region.  But the
simple criterion to determine that given vector is close to some
eigenvector exists only for hermitian matrices (see
sec.\ref{definition}).

Thus for the successful analysis of the asymptotic behavior of
anomalous dimensions two ingredients --
the hermiticity of the mixing matrix and the true choice of test
vector -- are essential. It is not evident that first condition
can be satisfied at all. But for the model under consideration
one can choose the scalar product in a such way that a mixing matrix
will be hermitian~\cite{WP,WK}. Some arguments in favor that it
can be done in a general case
will be given in  Sec.~\ref{conclusion}. As to
the choice of the trial vector this will be discussed below.

Henceforth, taking in mind the considerations given above, we shall
carry out the analysis of the asymptotic behavior of
anomalous dimensions for the whole class of the symmetric
and traceless operators in the scalar
$ \phi^{4} $ theory.

The paper is organized in the following way:
in sec.\ref{definition} we shall introduce  notations,
derive some formulae and give the exact formulation of the
problem; the sec.~\ref{sproof}
is devoted to the proof of the theorem about asymptotic behavior
of  anomalous  dimension, which is the main result of this
paper; in the last section we discuss the obtained results.

\section{Preliminary remarks}
\label{definition}
It had been shown
In the papers \cite{WP,WK}
that the problem
of  calculation of  anomalous dimensions of the
traceless and symmetric composite operators in the scalar
$  \phi^4 $
theory in the one--loop approximation
is equivalent to the eigenvalue problem for the
hermitian operator
${\rm H}$
acting on a Fock space
${\cal H}$:
\begin{equation}
{\rm H }=\frac{1}{2}
\sum_{n=0}^{\infty}\frac{1}{n+1} h_n^\dagger h_n= \frac{1}{2}
\sum_{n=0}^{\infty}\frac{1}{n+1} \sum_{i=0}^{n} a_i^\dagger
a_{n-i}^\dagger \sum_{j=0}^{n} a_j a_{n-j}.
\label{oper}
\end{equation}
Here $a_i^\dagger, a_i$  are the creation and annihilation
operators with the standard commutation relations
 $ [a_i,a_k^\dagger]=\delta_{ik}$.
The eigenvalues of
$ {\rm H} $ and the anomalous dimensions of composite operators
are simply related:
$\gamma_{an}=\epsilon/3\cdot\lambda+O(\epsilon^2)$.
There is also a one to one correspondence between the eigenvectors of
$ {\rm H} $ and the multiplicatively renormalized composite
operators~\cite{WK}.

It can be easily shown that
$  {\rm H} $ commutes with the  operator of particles number
$  {\rm N} $  and with the generators of
$ SL(2,C) $ group $\Sl$,
$ \Sp{} $,
$ \Sm{} $:
\begin{equation}
[\Sm{},{\rm S}] =  \Sm{} \ ,\
[\Sp{},{\rm S}] = - \Sp{} \ ,\
[\Sp{},\Sm{}] = 2{\rm S}.
\label{Sl2}
\end{equation}
They can be written as:
\begin{equation}
N = \sum_{j=0}^\infty  a_j^\dagger\ a_j,  \ \ \
 \Sl = \sum_{j=0}^\infty (j+1/2)\cdot a_{j}^\dagger\ a_j,
\label{number}
\end{equation}
\begin{equation}
\Sm{} = \sum_{j=0}^\infty (j+1)\cdot a_j^\dagger\ a_{j+1}, \ \ \
 \Sp{} = -\sum_{j=0}^\infty (j+1)\cdot a_{j+1}^\dagger\ a_j.
\label{sl2}
\end{equation}

Further,
due to commutativity of
$ {\rm H} $ with
$ SL(2,C) $ generators, each of the  subspaces
$  {\cal H}^{l}_{n}$
and  $ \bar {\cal  H}^{l}_{n}\in {\cal H}^{l}_{n}$
  ($n,l=0,\ldots,\infty $ ):
\begin{equation}
  {\cal  H}^{l}_{n}=\{ \psi\in {\cal H}|N\psi=n\psi\>,
\Sl\psi=(l+n/2)\psi\}, \>\>\>
\bar {\cal H}^{l}_{n}=\{ \psi\in {\cal H^{l}_{n}}|\Sm{}\psi=0 \}
\label{Hn}
\end{equation}
are invariant subspaces of the operator
$ {\rm H} $.
Since every eigenvector from
$ {\cal H}^{l}_{n} $ which is orthogonal to
$ \bar {\cal H}^{l}_{n} $  has the form~\cite{DM}:
$$
\vct{\psi}=\sum_{k}c_{k}\Sp{k}\vct{\psi_{\lambda}},\>\>\>\>
\vct{\psi_{\lambda}}\in\bar {\cal H}^{l}_{n},
$$
to obtain all spectrum of the operator
$ {\rm H} $  it
is sufficient to solve the eigenvalue problem for $ {\rm H} $
on each $  \bar {\cal H}^{l}_{n}$ separately.

Moreover, there exists a large subspace of the eigenvectors with
zero eigenvalues in each  $  \bar {\cal H}^{l}_{n}$.
They have been completely described in ref.~\cite{WK} and
will not be considered here.

As to nonzero eigenvalues, although at finite
$ l $ the spectrum of
$ {\rm H} $ has a very complicated structure (the numerical
results for particular values of
$ n $ and
$ l $  are given in refs.~\cite{WK,Kh}),
at large  $ l$  as it will be shown below  considerable
simplifications take place.

The main result of the present work
can be formulated in the form of the following theorem:
\newtheorem{th}{Theorem}
\begin{th}
Let eigenvectors
$ \psi_1\in \bar {\cal H}^r_n $
 and
$ \psi_2\in \bar {\cal
H}^s_m $ ($\psi_{1}\neq\psi_{2}$)
of operator
$ {\rm H} $
have the eigenvalues $
\lambda_1 $  and  $ \lambda_2 $ correspondingly.
Then there exists
a number $ L $  such,
that for every $ l\geq L$,  there exists
eigenvector $ \psi^l\in \bar {\cal H}^l_{(m+n)} $ with the
eigenvalue $ \lambda_l $ such that
\begin{equation}
|\lambda_l-\lambda_1-\lambda_2|\leq C\sqrt{\ln{l}}/l,
\label{eqth}
\end{equation}
where
$ C $ is some constant independent of l.
In the case when
$ \psi_{1}=\psi_{2} $ the same inequality holds only for
even
$ l\geq L $.
 \end{th}

The proof is based on the simple observation. Since any of
subspaces $  \bar {\cal H}^{l}_{n}$
 has a finite dimension, operator
$  {\rm H} $ restricted on $  \bar {\cal H}^l_n $ has only pointlike
spectrum. In this case
it can be easily shown  that if
there is a vector
$ \psi $, for which the condition
\begin{equation}
||({\rm H}-{\tilde \lambda})\psi||\leq\epsilon||\psi||
\label{condit}
\end{equation}
is fulfilled, then there exists the eigenvector
$ \psi_{\lambda} $
($ {\rm H}\psi_{\lambda}=\lambda\psi_{\lambda} $), such that
$ |\lambda-{\tilde \lambda}|\leq\epsilon$.
Indeed, expanding  a vector
$ \psi $  in the basis  of the eigenvectors of
$ {\rm H} $
$\psi=\sum_{k} c_{k}\psi_{k}   $ we obtain:
$$
\epsilon||\psi||\geq||({\rm H}-{\tilde \lambda})\psi||
=(\sum_{k}(\lambda_{k}-{\tilde \lambda})^{2}c_{k}^{2})^{1/2}\geq
\min_{k}|\lambda_{k}-{\tilde \lambda}|\cdot ||\psi||.
$$
So, to prove the theorem it is
sufficient to find out in the each subspace
$  \bar {\cal H}^{l}_{n+m} $
a vector which satisfies the
corresponding inequality.  Note, that for a nonhermitian matrix these
arguments are not applicable.

Before to proceed to the proof we give another formulation of
the eigenvalue problem for the operator
$ {\rm H} $. Let us note that there exists the one
to one correspondence between the vectors from
$ {\cal H}^{l}_{n} $ and the symmetric homogeneous
polynomials degree of
$ l $ of
$ n $ variables:
\begin{equation}
|{\Psi}>=\sum_{\{j_i\}}c_{j_1,\ldots,j_n}a^\dagger_{j_1}\ldots
a^\dagger_{j_n} \vac\to
{\psi}(z_1,\ldots,z_n)=
\sum_{\{j_i\}}c_{j_1,\ldots,j_n}z_1^{j_1}\ldots
z_n^{j_n},
\end{equation}
the coefficient
$ c_{j_1,\ldots,j_n}  $ being assumed totally symmetric.
It is evident that this mapping can be continued to all space.

The operators
$ \Sl $,
$ \Sp{}$,
$ \Sm{} $ in the
$ n $~-~particles sector take the form~\cite{DM}:
\begin{equation} \label{comm}
\Sl=
\sum_{i=1}^{n}(z_i \partial_{z_i}+1/2), \ \  \ \ \ \ \
\Sm{}= \sum_{i=1}^{n}
\partial_{z_i}, \  \   \ \ \ \ \
\Sp{}=
-\sum_{i=1}^{n}(z_i^2 \partial_{z_i} +
z_i).
\end{equation}
The operator
$ {\rm H} $ in its turn can be represented as the sum of the
two--particle hamiltonians:
\begin{equation}
{\rm H}=\sum_{i<k}^{n}{\rm H}(z_{i},z_{k}),
\label{opH}
\end{equation}
\vskip -0.6cm
where
\vskip -0.7cm
\begin{equation} {\rm
H}(z_{i},z_{k}) {\psi}(z_1,...,z_n) = \int_0^1 \rm{d}\alpha
{\psi}(z_1,...,
\alpha z_i + (1-\alpha)z_k,...,
\alpha z_i + (1-\alpha)z_k,...,z_n).
\label{acth}
\end{equation}
It should be stressed that not only
$ {\rm H} $, but every ${\rm H}(z_{i},z_{k})$ commutes with
$ \Sl, \Sp{}, \Sm{} $.

For further calculations it is very convenient to
put into correspondence to every
function of
$ n $ variables  the another one by the
following formula~\cite{DM}:
\begin{equation}
{\psi}(z_1,\ldots,z_n)
=\sum_{\{j_i\}}c_{j_1,\ldots,j_n}z_1^{j_1}\ldots
z_n^{j_n}\to
\phi
(z_1,
\ldots,z_n)=\sum_{\{j_i\}}
 ({j_1}!\cdots{j_n}!)^{-1}
c_{j_1,\ldots,j_n}z_1^{j_1}\ldots
z_n^{j_n}.
  \label{connection}
\end{equation}
The function
$ \psi $
 can be expressed in terms of $ \phi $
in the compact form:
\begin{equation}
{\psi}(z_1,\ldots,z_n)=
{\phi}(\partial_{x_1},\ldots,\partial_{x_n})
\prod_{i=1}^{n}\frac{1}{(1-x_{i}z_{i})}{\Bigl |}_{x_{1}=\ldots=x_{n}=0}
\label{1x}
\end{equation}
Then one obtains the following expression for the scalar
product for two vectors from
$ {\cal H}^{l}_{n} $:
\begin{equation}
<\psi_{1}|\psi_{2}>_{\cal H} =
n!\cdot{\phi}(\partial_{z_1},\ldots,\partial_{z_n})
{\psi}(z_1,\ldots,z_n)|_{z_{1}=\ldots=z_{n}=0}.
\label{norm1}
\end{equation}
It is easy now to check that
 the operators
$ \Sl,\ \Sp{},\ \Sm{},\ {\rm H} $ on the space of the
"conjugated" functions $ \phi(z_{1},\ldots,z_{n}) $ look as:
\begin{equation} \label{comm1}
\Sl=
\sum_{i=1}^{n}(z_i {\partial}_{z_i}+1/2), \ \  \ \ \ \ \
\Sp{}= \sum_{i=1}^{n}{ z_i}, \  \   \ \ \ \ \
\Sm{}=
-\sum_{i=1}^{n}(z_i {\partial}^{2}_{ z_i} +
{\partial}_{ z_i}),
\end{equation}
\begin{equation}
H {\phi}(z_1,...,z_n) =
\sum_{i<k} \int_0^1 \rm{d}\alpha
{\phi}(z_1,...,
(1-\alpha)(z_i+z_k),...,
\alpha (z_i + z_k),...,z_n).
\label{acth1}
\end{equation}

Up to now we assume the functions
${\psi}(z_1,\ldots,z_n)   $  to be totally symmetric.
But in the following we shall deal with  nonsymmetric
functions as well. To treat them on equal footing
it is useful to enlarge the
region of the definition of the operators
$ \Sl,\ \Sp{},\ \Sm{},\ {\rm H} $ up to the space of all polynomial
functions
  ${\cal B}=\bigoplus_{n,l=0}^{\infty}{\cal B}^{l}_{n}$,
where
$  {\cal B}^{l}_{n} $ is the linear space of the homogeneous
polynomials of degree
$ l $ of
$ n $ variables with the scalar product given by
eq.~(\ref{norm1}).
Then the Fock space
$ {\cal H} $ will be isomorphic to the subspace of the symmetric
functions of
$ {\cal B} $; and the subspace
$ \bar {\cal H}_{n}^{l} $~---~to the subspace of the symmetric
homogeneous translational invariant polynomials of degree $ l $
of $ n $ variables ${\hat{\cal B}}_{n}^{l}\in {\cal B}_{n}^{l} $.

\renewcommand{\theequation}{\thesubsection.\arabic{equation}}
\setcounter{equation}{0}
\section{Proof of Theorem}
\label{sproof}
\subsection{Part I}

Let us consider two eigenvectors of
$ {\rm H} $:
$ \psi_1\in \bar {\cal H}^r_n $ and  $ \psi_2\in \bar {\cal H}^s_m $
($ {\rm H}\psi_{1(2)}=\lambda_{1(2)}\psi_{1(2)} $ ); and let
$ \psin{1}{x}{n} $ and
$ \psin{2}{y}{m} $  are the symmetric translation--invariant
homogeneous polynomials corresponding to them, of degree
$ r $ and
$  s $ respectively. To prove the theorem it is enough to pick out in the
subspace
$\hat{\cal B}^{l}_{n+m} $ (or, the same, in the
$ \bar{\cal H}^{l}_{n+m} $) the function, for which the
inequality~(\ref{condit})  holds.

Let us consider the following function (nonsymmetric yet):
\begin{equation}
\PsiN{x}{y}=\sum_{k=0}^{l} c_{k}
(\mbox{\bf Ad}^{k}\Sp{})\Psin{1}{x}
(\mbox{\bf Ad}^{l-k}\Sp{})\Psin{2}{y}, \label{function}
\end{equation}
where $ c_k=(-1)^k\bin{l}{k}\cdot\bin{l+A++B}{k+A}  $
($\bin{l}{k}$ is the binomial coefficient);
\ \ \ $A=n+2r-1$;\ \ \
$ B=m+2s-1 $; \\  $ \mbox{\bf Ad}\Sp{}=[\Sp{},\cdot]$;\  and
for the brevity we used notations $\PsiN{x}{y}$\ \
for  \ $\psiN{}{x}{n}{y}{m}$, \ \ and
$\Psin{1(2)}{x}$ \ for
\  $\psin{1(2)}{x}{n(m)}$.

Using eq.~(\ref{Sl2}) it is straightforward to check that function
$ \psi $ given by eq.~(\ref{function}) is translational
invariant, i.e.
$ \Sm{}\PsiN{x}{y}=0 $.

To construct the test function we symmetrize
$ \PsiN{x}{y}$ over all
$ x $ and
$ y $:
\begin{equation}
\PSS{x}{y}=Sym_{\{x,y\}}\PsiN{x}{y}=
\frac{n!m!}{(n+m)!} \sum_{k=0}^{m} \sum_
{\stackrel{{\scriptstyle
\{i_{1}<\ldots<i_{k}\}}}{\{j_{1}<\ldots<j_{k}\}}}
\psi_{(j_{1}\ldots j_{k})}^{(i_{1}\ldots i_{k})}(x,y),
\label{symmetr}
\end{equation}
where
$
\psi_{(j_{1}\ldots j_{k})}^{(i_{1}\ldots i_{k})}(x,y)
$
is obtained from
$  \PsiN{x}{y}$ by  interchanging
$ x_{i_{1}} \leftrightarrow y_{j_{1}} $, and so on.
Also, without loss of generality, hereafter we take
$ n\geq m $. ( For the cases
$ m=1$ and  $n=1(2)$  the expression for the
$ \PSS{x}{y} $ yields the exact eigenfunctions, so one might hope that
it will be a good approximation  in  other cases also.)

The corresponding expression for the "conjugate"
function
$ \PhiN{x}{y} $   looks more simple.
Really, taking into account that
$ \Sp{}\phi(....)=(x_{1}+\cdots+x_{{n}})\phi(....) $ one
obtains:
\begin{eqnarray}
&&\PhiN{x}{y}=
\sum_{k=0}^{l}c_{k}(x_{1}+\cdots+x_{n})^{k}
(y_{1}+\cdots+y_{m})^{l-k}\Phin{1}{x}\Phin{2}{y}=\nonumber \\
&&=\Phin{1}{x}\Phin{2}{y}
K(a,b)
\exp{a(\sum_{i=1}^{n}x_{i})}
\exp{b(\sum_{j=1}^{m}x_{j})}
{\Bigl |}_{a=b=0},
\label{phifun}
\end{eqnarray}
where
\begin{equation}
 K(a,b)\equiv\sum_{k=0}^{l}c_{k} \partial_{a}^{k}\partial_{b}^{(l-k)}.
\label{K}
\end{equation}

Then, with the help of eqs.~(\ref{1x}),(\ref{phifun})
the following representation for
the function $ \PsiN{x}{y} $ can be  derived:
\begin{equation}
\PsiN{x}{y}=
\Phin{1}{{\partial_{\xi}}}
\Phin{2}{{\partial_{\eta}}}
K(a,b)
\prod_{i,j}\frac{1}{(1-x_{i}(a+\xi_{i}))}
\frac{1}{(1-y_{i}(b+\eta_{i}))}\Bigl|_{(a,b,\xi_{i},\eta_{j})=0}.
\label{represent}
\end{equation}

Now we are going to show that the inequality~(\ref{condit}) with
$\lambda=\lambda_{1}+\lambda_{2}$ and
$ \epsilon=C\sqrt{\ln{l}}/l $ holds for the function
$ \PSS{x}{y} $. As it was mentioned before this is
sufficient to the prove the theorem. Our first task is
the calculation of the norm of the  function $ \PSS{x}{y} $.
To be more precise, in the rest of this subsection we obtain the
estimate from below of the norm of $ \PSS{x}{y} $  for large
values of
$ l $.

Using eqs.~(\ref{norm1}),(\ref{represent}) and taking into
account that
$ \psi_{S}^{p}(x,y) $ is totally symmetric, one gets:
\begin{equation}
||\psi^{l}_{S}||^{2}=
n!m!\sum_{k=0}^{m}
\sum_
{\stackrel{{\scriptstyle
\{i_{1}<\ldots<i_{k}\}}}{\{j_{1}<\ldots<j_{k}\}}}
\PhiN{{\partial_x}}{{\partial_y}}
\psi_{(j_{1}\ldots j_{k})}^{(i_{1}\ldots
i_{k})}(x,y)=
{n!m!}\sum_{k=0}^{m}\bin{n}{k}\bin{m}{k}A^{(k)}_{l}.
\label{scprod}
\end{equation}
The coefficients
$ A^{(k)}_{l} $ are given by the formula:
\begin{equation}
A^{(k)}_{l}=\Np
\PhiN{{\partial_x}}{{\partial_y}}
\psi^{(1\ldots k)}_{(1\ldots k)}(x,y),
=\Np\phi_{1}(\partial_{{ x}_{i}})
\phi_{2}(\partial_{{ y}_{i}})
K({ a},{ b})
\psi^{(1\ldots k)}_{(1\ldots k)}({ x},{ y}+{ b}-{ a}),
\label{defA}
\end{equation}
where the  symbol
$ \Np $ denotes that in the end of  calculation all arguments must be
set to zero.

The substitution of (\ref{phifun}),~(\ref{represent})
into (\ref{defA}) yields :
\begin{eqnarray}
A^{(k)}_{l}&=& \Np\Phin{1}{{\partial_{x}}}
\Phin{2}{{\partial_{y}}}
\Phin{1}{{\partial_{\bar x}}}
\Phin{2}{{\partial_{\bar y}}}
K(a,b)K({\bar a},{\bar b})\Biggl [
\prod_{i=k+1}^{m}
\frac{1}{(1-({\bar  y}_{i}+{\bar  b}-{\bar  a})({y}_{i}+{b}))}
\nonumber \\
\lefteqn{\prod_{i=1}^{k}
\frac{1}{(1-({\bar  y}_{i}+{\bar b}-{\bar a})({x}_{i}+{a}))
(1-{\bar x}_{i}({y}_{i}+{b}))}
\prod_{i=k+1}^{n}
\frac{1}{(1-{\bar x}_{i}({x}_{i}+{b}))} \Biggr ].}
\label{defS}
\end{eqnarray}
The expression in the square brackets of~(\ref{defS})
depends only  on the difference
$ {\bar b}-{\bar  a} $, hence
$$
K({\bar  a},{\bar  b})\Biggl [....\Biggr]_{{\bar  a},{\bar  b}..=0}=
\Bigl (\sum_{k}(-1)^{k}c_{k} \Bigr ) \partial^{l}_{{\bar  b}}
\Biggl [....\Biggr]_{{\bar  a},{\bar  b}..=0}.
$$
In the resulting expression the dependence on
$ {\bar b} $ can be factorized after the appropriate rescaling of the
variables
$ x,y,{\bar x},{\bar y},{a}, {b} $.
At last, taking advantage of eq.~(\ref{1x}) and
remembering
 that the function
$ \psi_{1},\ \psi_{2} $ (but not
$ \phi(...) $) are  translation invariant, one gets:
\begin{equation}
A^{(k)}_{p}= Z \Np
\phi_{1}(\partial_{{x}_{i}})
\phi_{2}(\partial_{{y}_{i}})K({a},{b})F_{k}(a,b,x,y),
\label{DFS}
\end{equation}
where
$ Z=l!\sum_{k=0}^l c_k (-1)^k=l!\bin{2l+A+B}{l+A}$ and
\begin{eqnarray}
&&F_{k}(a,b,x,y)=
\Biggl [
\psi_{1}({y}_{1},\ldots,{y}_{k},{x}_{k+1}+{a}-{b},
\ldots,{x}_{n}+{a}-{b})
\prod_{i=1}^{k}\frac{1}{{1-{x}_{i}-{a}}}
\prod_{i=k+1}^{m}\frac{1}{{1-{y}_{i}-{b}}}
\nonumber \\
\lefteqn{\mbox{\hskip 2cm}(1-{a})^{-s}
\psi_{2}(\frac{{x}_{1}}{1-{x}_{1}-{a}},
\ldots,
\frac{{x}_{k}}{1-{x}_{k}-{a}},
\frac{{y}_{k+1}+{b}-{a}}{1-{y}_{k+1}-{b}},\ldots,
\frac{{y}_{m}+{b}-{a}}{1-{y_{m}-{b}}}) \Biggr ].}
\label{defS1}
\end{eqnarray}
It is easy to understand that after the differentiation with
respect to $ x_{i}, y_{j} $ the resulting expression will have form:
\begin{equation}
A^{(k)}_{l}=Z\Np K(a,b)\sum_{n_{1},n_{2},n_{3}} C^{k}_{n_{1},n_{2},n_{3}}
\frac{(a-b)^{n_{1}}}{(1-a)^{n_{2}}(1-b)^{n_{3}}}=
Z\sum_{n_{1},n_{2},n_{3}} C^{k}_{n_{1},n_{2},n_{3}}
A^{k}_{n_{1},n_{2},n_{3}}(l),
\label{nnn}
\end{equation}
the summation over $n_{1},\ n_{2},\ n_{3}$ being carried out in the limits,
which as well as the coefficients $ C^{k}_{n_{1},n_{2},n_{3}}$  are
independent from the parameter
$ l $. Thus, all dependence on
$ l $ of $ A^{(k)}_{l} $, except for the trivial factor
$ Z $, is contained in the coefficients $
A^{k}_{n_{1},n_{2},n_{3}}$.

Our further strategy is the following:
first of all we shall obtain
the result for the quantity
$ A^{(0)}_{l} $ and for
$ A^{(m)}_{l} $.
(These terms gives  the main contributions to the
norm of the vector
$ \psi_{S}^{l} $.) Then, we shall show (it will be done in Appendix A)
that for all other
$ A^{k}_{l} $ ($ 1\leq k \leq m-1  $)
for which we are not able to get the exact result,
the ratio
$A^{k}_{l}/A^{0}_{l} $
tends to zero as
$ 1/l^{2} $ at least.

To calculate
$ A^{(0)}_{l} $ it is sufficient to note that
 the expression for
$ F_{0}(a,b,x,y) $ (eq.~(\ref{defS1}))
 after the appropriate shift of the arguments
in  the functions
$ \psi_{1}  $ and
$ \psi_{2} $ reads:
\begin{equation}
\Biggl [
\psi_{1}({x}_{1},\ldots,{x}_{n})
\prod_{i=1}^{m}\frac{1}{{1-{y}_{i}-{b}}}
 (1-{b})^{-s}
\psi_{2}(\frac{{y}_{1}}{1-{y}_{1}-{b}},
\ldots,
\frac{{y}_{m}}{1-{y}_{m}-{b}})
 \Biggr ].
\label{as1}
\end{equation}
Then  carrying out the differentiation with respect to
$ x_{i}, y_{j} $ in eq.~(\ref{DFS}) one obtains:
\begin{equation}
A^{(0)}_{l}=Z(m!n!)^{-1}||\psi_{1}||^{2}||\psi_{2}||^{2}\Np
K(a,b)(1-b)^{-(2s+m)}
= (m!n!)^{-1}{\cal A}(l),
\end{equation}
where
\begin{equation}
{\cal A}(l)={(2l+A+B)!}\frac{l!(l+A+B)!}{{(l+A)!(l+B)!}}
\frac{||\psi_{1}||^{2}||\psi_{2}||^{2}}{A!B!}.
\label{A0}
\end{equation}
The evaluation of
$ A^{m}_{l} $ in the case when
$ n=m=k $ differs from the considered above only
in the interchange of variables
$ x,a\leftrightarrow y,b $ in~(\ref{as1}).
Then the  straightforward calculations yield:
\begin{equation}
A^{(m)}_{l}=Z|<\psi_{1}|\psi_{2}>|^{2}\Np
K(a,b)(1-a)^{-(2s+m)}=(-1)^{l}A^{(0)}_{l}\delta_{\psi_{1}\psi_{2}}.
\label{Am}
\end{equation}
Here, we take into account that
$ \psi_{1}, \ \psi_{2} $ are the eigenfunctions of the self-adjoint
operator.
Thus, when
$ l $ is odd and
$\psi_{1}=\psi_{2} $  these two contributions
($ A^{(0)}  $ and
$ A^{(m)} $)
cancel each other.
But, as one can easily see from eq.~(\ref{function}),
the function
$ \PsiN{x}{y} $ is identically equal to zero in this case.

In the case when $k=m$ and
$ m<n $  eq.~(\ref{nnn}) for
$ A^{(m)}_{l} $ reads:
\begin{equation}
 A^{(m)}_{l}=Z\Np
 K(a,b)\sum_{z=s}^{r}c_{z}\frac{(a-b)^{z-s}}{(1-a)^{s+z+m}}=
Z\Np K(a,b)\sum_{k=0}^{r-s}{\tilde
c}_{z}\frac{(1-b)^{k}}{(1-a)^{2s+m+k}}.
\label{amn}
\end{equation}
After some algebra one obtains:
\begin{equation}
 A^{(m)}_{l}=Z
(l+A+B)!\sum_{k=0}^{r-s}\sum_{i=0}^{k}{\tilde
c}_{k,i} \frac{l!(l+B+k-i)!}{(l-i)!(l+A-i)!}\leq c\frac{A^{(0)}_{l}}{l^{n-m}}
\leq c\frac{A^{(0)}_{l}}{l}.
\label{am1}
\end{equation}
The similar calculations (see Appendix~A for details)
in the case of
$ 0< k< m $ give:
\begin{equation}
 |A^{(k)}_{l}|\leq \alpha_{k}A^{(0)}_{l}/l^{2},
\label{am2}
\end{equation}
where
$ \alpha_{k} $ are some constants.
Then, taking into account (\ref{A0}), (\ref{Am}) (\ref{am1}), (\ref{am2}),
one obtains:
\begin{equation}
||\psi_{S}^{l}||^{2}=(1+(-1)^{l}\delta_{\psi_{1},\psi_{2}})
{\cal A}(l)(1+O(1/l)).
\label{resnor}
\end{equation}

\setcounter{equation}{0}
\subsection{Part II}

To complete the proof of the theorem it is remained  to obtain the
following inequality:
\begin{equation}
\eps=||\hd\psi^{l}_{S}||^{2}=
||({\rm H}-\lambda_{1}-\lambda_{2})\psi^{l}_{S}||^{2}\leq
C\ln{l}/l^{2} {\cal A}(l)
\label{eps1}
\end{equation}
for
$ l\geq l_{0} $.

To do this let us, first of all, substitute the
expression~(\ref{symmetr}) for $ \psi^{l}_{S}$ in~(\ref{eps1}).
Then, taking into account the invariance of the operator
$ {\rm H} $ under any transposition of its arguments (see~
eq.(\ref{opH}))  one gets:
\begin{equation}
(\eps)^{1/2}=\frac{n!m!}{(n+m)!}||
\sum_{k=0}^{m}
\sum_
{\stackrel{{\scriptstyle
\{i_{1}<\ldots<i_{k}\}}}{\{j_{1}<\ldots<j_{k}\}}}
\hd\psi_{(j_{1}\ldots j_{k})}^{(i_{1}\ldots i_{k})}(x,y)||\leq
||\hd\psi^{l}(x,y)||.
\label{eps2}
\end{equation}
Further, from the eqs.~(\ref{opH}),~(\ref{function}) the equality
\begin{equation}
\hd\psi^{l}(x,y)=
\sum_{i,k}H(x_{i},y_{k})\psi^{l}(x,y)
\end{equation}
immediately follows.  At last,  taking into consideration that
$$ ||H(x_{i},y_{k})\psi^{l}(x,y)||=||H(x_{1},y_{1})\psi^{l}(x,y)|| $$
\vskip -0.3cm
and
\vskip -0.3cm
$$ ||H(x_{1},y_{1})\psi^{l}(x,y)||^{2}=
 <\psi^{l}(x,y)H(x_{1},y_{1})\psi^{l}(x,y)>$$
we  obtain the following estimate of
$ \eps $:
\begin{equation}
\eps\leq
(mn)^{2}<\psi^{l}(x,y)H(x_{1},y_{1})\psi^{l}(x,y)>.
\label{eps3}
\end{equation}

Our next purpose is to obtain the expression
for the matrix element in~(\ref{eps3})
similar to that  for
$ A^{k}_{l} $ (eq.~(\ref{nnn})).
This matrix element, with the help of
formulae~(\ref{phifun}),(\ref{represent}), can be
represented in the following form:
\begin{eqnarray}
\lefteqn{<\psi^{l}(x,y)H(x_{1},y_{1})\psi^{l}(x,y)>=Z
(n+m)!
K(a,b)
\Phin{1}{\partial_{x}}
\Phin{2}{\partial_{y}}
\prod_{i=2}^{n}\frac{1}{1-ax_{i}}
\prod_{i=2}^{m}\frac{1}{1-by_{i}}} \nonumber \\
&& \int_{0}^{1} ds
\frac{1}{1-a\theta(s)}
\frac{1}{1-b\theta(s)}
\psi_{1}\Bigl (\frac{\theta(s)}{1-a\theta(s)},\frac{1}{1-ax_{i}}\Bigl )
\psi_{2}\Bigl (\frac{\theta(s)}{1-b\theta(s)},\frac{1}{1-by_{i}}\Bigl )
\Bigl |_{\stackrel{{\scriptstyle x=0,y=1}}{a=b=0}},
\label{rH}
\end{eqnarray}
where
$ \theta(s)=sx_{1}+(1-s)y_{1} $.
After the differentiation with respect to
$ x_{i},\ y_{j}, i,j>1 $ one gets:
\begin{equation}
<.....>=
\sum_{n_{1}=0}^{r}
\sum_{n_{2}=n_{1}}^{r}
\sum_{m_{1}=0}^{s}
\sum_{m_{2}=0}^{s}
\sum_{m_{3}=0}^{s-m_{1}}
c^{n_{1}n_{2}}_{m_{1},m_{2},m_{3}}
{\tilde a}^{n_{1}n_{2}}_{m_{1},m_{2},m_{3}}(l),
\label{rH0}
\end{equation}
where
\begin{equation}
{\tilde a}^{n_{1}n_{2}}_{m_{1},m_{2},m_{3}}(l)=
Z K(a,b)\Biggl [
\frac{a^{n_{2}-n_{1}}}{(1-b)^{\bt-m_{2}}}
\partial_{x}^{n_{1}}
\partial_{y}^{m_{1}}
\int_{0}^{1}ds
\frac{\theta^{n_{2}+m_{2}}}{(1-a\theta)^{n_{2}+1}(1-b\theta)^{m_{2}+1}}
\Biggr ]_{\stackrel{{\scriptstyle x=0,y=1}}{a=b=0}}
\label{rH1}
\end{equation}
and
$ \bt=s+m_{3}+m-1 $. Note that the coefficients
$c^{n_{1}n_{2}}_{m_{1},m_{2},m_{3}}$in eq.~(\ref{rH0})
 do not depend on
$ l $.

Before  applying the operator
$ K(a,b)=\sum_{k=0}^{l}c_{k}\partial_{a}^{k}\partial_{b}^{l-k} $
to the expression in the square brackets
it is convenient to rewrite the latter
in more suitable for this purpose form:
\begin{equation}
\Biggr [....  \Biggl]=(n_{2}!m_{2}!)^{-1}
\frac{a^{n_{2}-n_{1}}}{(1-b)^{\bt-m_{2}}}
\int_{0}^{1}ds
s^{m_{1}}(1-s)^{n_{1}}
\partial_{a}^{n_{2}}
\partial_{b}^{m_{2}}
\partial_{s}^{(m_{1}+n_{1})}
\frac{1}{(1-as)(1-bs)}.
\label{sqb}
\end{equation}
Now  all differentiations with respect to
$ a $ and
$ b $ can be easily fulfilled:
$$
\partial_{a}^{k}
a^{n_{2}-n_{1}}
\partial_{a}^{n_{2}}
\frac{1}{(1-as)}\bigl |_{a=0}
=(k+n_{1})!s^{k+n_{1}}
\partial_{x}^{n_{2}-n_{1}}x^{k}\bigl |_{x=1},
$$
$$
\partial_{b}^{l-k}
\frac{1}{(1-b)^{\bt}}
\partial_{b}^{m_{2}}
\frac{1}{(1-sb)}=
s^{m_{2}}\frac{(l-k+\bt)!}{\Gamma(\bt-m_{2})}
\int_{0}^{1}d\al\al^{\bt-m_{2}-1}(1-\al)^{m_{2}}[\al+(1-\al)s]^{l-k}.
$$
At last,  after a representation of
the ratio of the factorials like
$ (k+n_{1})!/(k+A)!$ in the form
$ 1/\Gamma(A-n_{1}-1)=\int_{0}^{1}du u^{k+n_{1}}(1-u)^{A-n_{1}-1} $
the summation over
$ k $  becomes trivial and we obtain
the following expression for
$
{\tilde a}^{n_{1}n_{2}}_{m_{1},m_{2},m_{3}}(l)  $:
\vskip -0.3cm
\begin{equation}
{\tilde a}^{n_{1}n_{2}}_{m_{1},m_{2},m_{3}}(l)=
{\cal A}(l)
a^{n_{1}n_{2}}_{m_{1},m_{2},m_{3}}(l),
\label{newbb}
\end{equation}
\vskip -0.3cm
where
\begin{eqnarray}
\lefteqn{a^{n_{1}n_{2}}_{m_{1},m_{2},m_{3}}(l)=
\frac{1}{{\Gamma(A-n_{1})\Gamma(\bt-m_{2})\Gamma(B-\bt)}}
\partial_{x}^{n_{2}-n_{1}}
\int_{0}^{1}..\int_{0}^{1}
ds\, d\al\, du\, dv\,
u^{n_{1}}(1-u)^{A-n_{1}-1}s^{m_{1}}}
\nonumber \\
&&
(1-s)^{n_{1}}
\partial_{s}^{n_{1}+m_{1}}s^{n_{1}+m_{2}}
v^{\bt}(1-v)^{B-\bt-1}
\al^{\bt-m_{2}-1}
(1-\al)^{m_{2}}[v(\al+(1-\al)s)-sux]^{l}{\biggl |}_{x=1}.
\label{bb}
\end{eqnarray}
\vskip +0.2cm
Note, that when the arguments of
$ \Gamma $ -- functions become equal to
zero, the following evident changes
must be done:\ \
$ 1/\Gamma(A-n_{1})\int du (1-u)^{{A-n_{1}-1}}..\to \int du \delta(1-u) $
if
$ A=n_{1} $, and so on.
With the account of  eq.~(\ref{newbb})
  eq.~(\ref{rH1}) reads:
\begin{equation}
<\psi^{l}(x,y)H(x_{1},y_{1})\psi^{l}(x,y)>= {\cal A}(l)
\sum_{n_{1}=0}^{r}
\sum_{n_{2}=n_{1}}^{r}
\sum_{m_{1}=0}^{s}
\sum_{m_{2}=0}^{s}
\sum_{m_{3}=0}^{s-m_{1}}c^{n_{1}n_{2}}_{m_{1},m_{2},m_{3}}
                        a^{n_{1}n_{2}}_{m_{1},m_{2},m_{3}}(l).
\label{bb0}
\end{equation}
Thus to prove the inequality~(\ref{eps1}) for
$ \eps $ we should only  show that the coefficients
$ a^{n_{1}n_{2}}_{m_{1},m_{2},m_{3}}(l) $
  tend to zero not slowly than
$ \ln{l}/l^{2} $ at $ l\to \infty $.

First of all, let us consider the cases when at least one
$ \Gamma $  function in~(\ref{bb}) has zero argument.
(It is worth to remind that
$ A=2r+n-1 $,
$ B=2s+m-1 $; $n,\ m$ ($m\leq n$) are the numbers of variables and
$r, s$ -- degrees of the translational invariant polynomials
$ \psi_{1} $ and
$ \psi_{2} $  correspondingly).

\noindent
1.\ \ \
$ A=n_{1} $.
It is easy to understand that equality
$ A-n_{1}=0 $ is possible only when
$ n=1,\ r=0 $, and consequently  $ m=1,\ s=0 $.
 Then one immediately obtains that arguments of two other
$\Gamma $ functions are also zero, and the corresponding
integral~(see eq.~(\ref{bb}) and the note to it) is zero.

\noindent
2.\ \ \
$ \bt-m_{2}=0 $,
$ n>1 $ ($ 0<A-n_{1} $).
In this case one obtains
$ m=1,\ s=0 $ and
$ B-\bt=0 $.  Since two of  $\Gamma $ functions
have  arguments equal to zero, the integration over
$ v $ and
$ \al $ are removed. After this it is trivial to check
that
$ a(l) $ tends to zero as
$ 1/l^{2} $ at
$ l\to\infty $.

\noindent
3.\ \ \
$ B-\bt=0 $ and $ 1<m\leq n $.
Evidently, this is possible only when
$ m_{1}=0 $ and
$ m_{3}=s $. Again one integration (over $ v $) is removed.
To calculate
$ a(l) $ first of all let us write
$\partial_{x}^{n_{2}-n_{1}}$
in~(\ref{bb}) as
$  u^{n_{2}-n_{1}}\partial_{u}^{n_{2}-n_{1}} $ and carry out
the integration by parts as over
$ u $ as well as over
$ s $. Note, that the boundary terms do not appear when
$ m_{1}=0 $. Then it is clear that integrand represent by itself
the product
of two functions, one of which, $ [(\al+(1-\al)s)-su]^{l}$,
is positive definite in the region of integration
and other is the sum of the monoms like
$ s^{i_{1}}(1-s)^{i_{2}}\al^{i_{2}}... $ with finite coefficients.
But since
$ 0\leq s,u,\al\leq 1 $ this sum can be limited by some constant
which is independent of
$ l $. Then, taking into account this remark, one obtains
 the following estimate of $a(l) $:
\begin{eqnarray}
\lefteqn{|a^{n_{1}n_{2}}_{m_{1},m_{2},m_{3}}(l)|\leq C
\int_{0}^{1}..\int_{0}^{1}ds\, d\al\,
du\,[(\al+(1-\al)s)-su]^{l}=} \\
&&=\frac{C}{(l+1)}\int_{0}^{1}ds\, du\,
\frac{(1-su)^{l+1}-s^{l+1}(1-u)^{l+1}}{1-s}=\frac{2C}{(l+1)(l+2)}
[\psi(l+2)-\psi(1)],  \nonumber
\label{bb1}
\end{eqnarray}
where
$ \psi(x)=\partial_{x}\ln{\Gamma(x)} $.

\noindent
4.\ \ \ At last, we consider the case when all arguments of
$ \Gamma $ functions in eq.~(\ref{bb}) are greater then zero.
As in the previous case
it is convenient to replace
$\partial_{x}^{n_{2}-n_{1}}$ with
$  u^{n_{2}-n_{1}}\partial_{u}^{n_{2}-n_{1}} $ and fulfil
the integration  over
$ u $ and $ s $ by parts.
But the boundary terms arise now with the integration
over
$ s $ at upper bound ($s=1$).
However, it is not hard to show that each of them  decreases as
$ 1/l^{2} $ at
$ l\to\infty $.  (All calculations practically repeat those given
in Appendix~A)

The last term to be calculated has the form:
\begin{equation}
I(l)=
\int\int\int
ds\, d\al\, du\, dv\,
A(s,\al,u,v)[v(\al+(1-\al)s)-su]^{l},
\end{equation}
where $A(s,\al,u,v)$ is some polynomial of variables
$s,\,  \al,\, u,\, v\,$, such that
$A(s,\al,u,v)<C$  in the domain
$0\leq s,\, \al,\, u,\, v\, \leq 1$.
Then for even
$ l $ one obtains:
\begin{equation}
I(l)\leq C
\int_{0}^{1}\ldots\int_{0}^{1}
ds\, d\al\, du\, dv\,
[v(\al+(1-\al)s)-su]^{l}.
\end{equation}
If
$ p $ is odd, let us divide the domain of the integration into two
regions~--~$ \Omega_{+} $ and $ \Omega_{-} $~---~ in which the function
$[v(\al+(1-\al)s)-su]$  is either positive or   negative.
It is straightforward to get that
$ \Omega_{+},\ \Omega_{-} $ are defined by the following conditions:
\begin{eqnarray}
\Omega_{+}:\ &&
(\ u\leq v) \ \mbox{\rm or } (v\leq u,\ s\leq v/u, \
s(u-v)/v(1-s)\leq \al\leq 1);  \nonumber \\
\Omega_{-}:\  &&
(v\leq u);\ \mbox{\rm and }( s\geq v/u \ \mbox{\rm or }
(s\leq v/u,\ 0\leq \al\leq s(u-v)/v(1-s))).  \nonumber
\end{eqnarray}
Then integral $I(l)$ is estimated as:
\begin{equation}
I(l)\leq C_{+}
\int_{\Omega_{+}}
ds\, d\al\, du\, dv\,
[v(\al+(1-\al)s)-su]^{l}
- C_{-}
\int_{\Omega_{-}}
ds\, d\al\, du\, dv\,
[v(\al+(1-\al)s)-su]^{l}.
\end{equation}
The evaluation of the corresponding integrals does not cause
any troubles and leads to the following result:
\begin{equation}
  I(l)\leq C \ln{l}/l^{2}.
\end{equation}

Then taking into account eqs.~(\ref{eps1}),(\ref{eps3}),(\ref{bb0})  and
(\ref{resnor}) one concludes that there exist such constants
$ L$  and
$ C $ that inequality
\begin{equation}
||({\rm H}-\lambda_{1}-\lambda_{2})\psi^{l}_{S}||^{2}\leq
C\ln{l}/l^{2}||\psi^{l}_{S}||^{2}
\label{res1}
\end{equation}
holds for all
$ l\geq L $.
This inequality, as it has been shown in  sec.~\ref{definition}
 guarantees the existence of
the eigenvector  of the operator
$ {\rm H} $ with the eigenvalue satisfying the
 eq.~(\ref{eqth}).~\rule[-1pt]{9pt}{9pt}

\section{Conclusion}
\label{conclusion}

The theorem proven in the previous section provides a number
of consequences for the spectrum of the operator
$ {\rm H} $:
\begin{itemize}
\item Every point of the spectrum is either a limit point of the
latter  or an exact eigenvalue of  infinite multiplicity;
\item Any finite sum of eigenvalues and limit points of the
spectrum is a limit point again.
\end{itemize}
These statements directly follow from the theorem.

Further, let us denote by
$ {\cal S}_{n} $
the spectrum of operator
$ {\rm H} $
restricted on
$ n $~--~particle sector of Fock space,
($N\psi=n\psi$) and by
$ {\bar{\cal S}}_{n} $
the set of the limit points of
$ {\cal S}_{n} $.
Then it is easy to see that the definite relations
("hierarchical structures")
between
$ {\cal S}_{n} $,
$ {\bar{\cal S}}_{n} $  ($n=2,\ldots\infty$) exist.
Indeed, let
$ \Sigma_{n} $ is the set of all possible sums of
$ s_{i_{1}}+s_{i_{2}}+\cdots+s_{i_{m}} $ type, where
$s_{i_{k}}\in{\cal S}_{k} $ and
$ {i_{1}}+\cdots+{i_{m}}\leq n $,
$ {i_{1}}\leq {i_{2}}\leq n \cdots\leq {i_{m}}$. Then
one can easily conclude that the following relations
$$ \Sigma_{n}\subset {\bar{\cal S}}_{n} $$
are valid.
For
$ n=2,3 $ the more strong equations
$ \Sigma_{n}={\bar{\cal S}}_{n} $ hold,
but the examination of this conjecture for a general
$ n $ requires  additional analysis.

So one   can see that a number of interesting properties are
specific to one--loop spectrum of anomalous dimensions of
composite operators in
$ \phi^{4} $ theory. Of course, two question arise: which
of these (one--loop) properties survive in a higher order of
the perturbation theory? And, in what extent they are conditioned
by the peculiarity of
$ \phi^{4} $ theory?

As to the first question we can only adduce some arguments in favor
of a nonperturbative character of the obtained results. In the
paper~\cite{Kh} the spectrum of critical exponents of the
$ N $--vector model in
$ 4-\epsilon $ dimensions was investigated to the second order in
$ \epsilon $. In this work it was obtained  that some one--loop
properties of the spectrum, a generic class of
degeneracies~\cite{WK,DM} in particular,  are lifted in two--loop
order. However, the results of the numerical analysis of critical
exponents carried out for the operators with number of fields
$ \leq 4 $ distinctly show that a limit points structure of the
spectrum is preserved.

The other evidence in favor of this hypothesis can be found in
the works of~K.~Lang and W.~R\"{u}hl~\cite{Ruehl}. They investigated
the spectrum of  critical exponents in the nonlinear
$ \sigma $ model in
$ 2<d<4 $ dimensions in  the first order of
$ 1/N $ expansion. The results for various classes of composite
operators~\cite{Ruehl} display the existence of a similar
limit points structures in this model for whole range $ 2<d<4 $.
Since the critical exponents should be consistent in
$ 1/N $ expansion for
$ \sigma $ model and the
$ 4-\epsilon $ expansion for
$ (\phi^{2})^{2} $ model one may expect this property of the
spectrum to hold to all orders in the
$ \epsilon $ in the latter.

To answer the second question it is useful to understand what
features of the model under consideration determine the
properties of operator
$ {\rm H} $ --- hermiticity, invariance under
$ SL(2,C) $ group, two-particle type of interaction ---
which were crucial for
the proof of the theorem. First two properties are closely related to
the conformal invariance of the
$ \phi^{4} $ model~\cite{Shafer}.
It can be shown that a two-particle form of operator
$ {\rm H} $ and the conformal invariance of a theory lead
to  hermiticity of
$ {\rm H} $ in the scalar product given by eq.~\ref{norm1}.
(The relation between the functions
$ \psi $ and
$ \phi $ in the  general case is given in ref.~\cite{Hep}.) Further,
it should be emphasized that the commutativity of
$ {\rm H} $ with
$ \Sl $ and
$ \Sp{} $ reflects two simple facts:\  \
1. Nontrivial mixing occurs (in $\phi^{4}$ theory) only between
operators with equal number of fields.\ \
2. The total derivative of a eigenoperator is an eigenoperator with
the same anomalous dimension as well.

But if operator
$ {\rm H} $  is  hermitian it must commute with the operator
conjugated to
$ \Sp{} $ as well. So the minimal group of invariance of
$ {\rm H} $
($SL(2,C)$ in our case) has three generators.

Thus, the method of  analysis of anomalous dimensions of
composite operators in the
$ l\to\infty $ limit presented here is not peculiar for
$ \phi^{4} $ theory only and can be applied to other theories,
which are conformal invariant at one--loop level.

\vspace {0,5cm}
\centerline{\bf Acknowledgments}
\vspace {0,5cm}

Authors are grateful to Dr.~S.~Kehrein and Dr.~A.A.~Bolokhov
for stimulating
discussions and critical remarks.

The work was supported by Grant  95-01-00569a of Russian Fond of Fundamental
Research.

\setcounter{equation}{0}
\setcounter{section}{0}
\renewcommand{\thesection}{{}}

\section{Appendix A}
\renewcommand{\thesection}{\Alph{section}}
\renewcommand{\theequation}{\thesection.\arabic{equation}}
\label{ApA}
In this appendix we calculate the quantities
$ A^{(k)}_{l} $  for
$ 1\leq k\leq m-1  $.
 Let us remind the representation for
$ A^{(k)}_{l} $  (see eq.~(\ref{nnn})):
\begin{equation}
A^{(k)}_{p}=Z\Np K(a,b)\sum_{n_{1},n_{2},n_{3}} C^{k}_{n_{1},n_{2},n_{3}}
\frac{(a-b)^{n_{1}}}{(1-a)^{n_{2}}(1-b)^{n_{3}}}.
\label{nnn1}
\end{equation}
The summation
over $n_{1},n_{2},n_{3}   $ is carried out in the following range:
$$
n_{1}=m_{1}+m_{2}; \ \ \ \ n_{2}=k+s+m_{1}+m_{3}; \ \ \
n_{3}=s+m-k+m_{2}-m_{3};
$$
\begin{equation}
 0\leq m_{1}\leq \min{[s,r]};\ \ 0\leq m_{2}\leq s;\ \ \
0\leq m_{3}\leq r-m_{1}.
\label{range}
\end{equation}
The more simple way to obtain these bounds from (\ref{defS1}) is to treat
$  (a-b),\  (1-a),\  (1-b) $ as independent variables.

In the course of  calculation  of the coefficient
$ A^{(k)}_{n_{1},n_{2},n_{3}}=\Np K(a,b)
(a-b)^{n_{1}}(1-a)^{-n_{2}}(1-b)^{-n_{3}}$
we shall not look after the
factors  independent of
$ l $.
Then taking advantage of the Feynman's formula for the product
${(1-a)^{-n_{2}}(1-b)^{-n_{3}}}$ one obtains:
\begin{equation}
A^{(k)}_{n_{1},n_{2},n_{3}}\sim \sum_{k=0}^{l}c_{k}
\partial_{a}^{k}\partial_{b}^{(l-k)}
\int_{0}^{1}dx x^{n_{2}-1}(1-x)^{n_{3}-1}\partial_{x}^{n_{1}}
\frac{1}{[1-ax-(1-x)b]^{B+1}}.
\label{777}
\end{equation}
(let us remind that $ B=2s+m-1, \ A=2n+r-1 $, and
$ m\leq n $).
After the integration by parts  in~(\ref{777})
(note, that in the case
$ r\leq s $ there are no
boundary terms)
and the differentiation with respect to
$ a,b $ and
$ x $ one has:
\begin{eqnarray}
\lefteqn{A^{(k)}_{n_{1},n_{2},n_{3}}\sim
\sum_{k=0}^{l}c_{k}(l+B)!\sum_{j=n_{3}-n_{1}-1}^{B-1} \alpha_{j}
\int_{0}^{1} dx x^{k}(1-x)^{l-k+j}=}\nonumber \\
&&(l+B+A)!
\sum_{j}\alpha_{j} \frac{(l+B)!}{(l+j+1)!}
\sum_{k=0}^{l}(-1)^{k}\bin{l}{k}
\frac{k!}{(k+A)!}\frac{(l-k-j)!}{(l-k+B)!}\sim \\
&&\sim\sum_{j=n_{3}-n_{1}-1}^{B-1} \alpha_{j}\frac{(l+B)!(l+B+A)!}{(l+j+1)!}
\int_{0}^{1}du dv u^{A-1}(1-v)^{j}v^{{B-j-1}}(u-v)^{l}.
\label{integ}
\end{eqnarray}
Here,
$ \al_j $ are some unessential constants.
Since
$ |(u-v)|\leq 1 $ every  integral in~(\ref{integ}) tends to zero at
$ l\to\infty $. For more precise estimate it is convenient
to divide the
domain of integration into two regions
($ u\leq v $ and
$ v\leq u $) and  to rescale the variables
$ v=ut $ ($u=vt$) in each of two integrals. Then replacing
all terms of
$ (1-ut)^{{\al}} $ type by unit one obtains the answer in the form
of the product of two beta functions.
Taking into account that $  s-r+m-k-1\leq j \leq B-1 $ and
collecting all necessary terms one gets that the
contribution from $ A^{(k)}_{n_{1},n_{2},n_{3}}  $ to the
$ A^{(k)}_{l} $ is of order
$ {\cal A}(l)/l^{2} $.

Thus we have obtained the required result for the case
$ r\leq s $. To do the same for the
$ s\leq r $, first of all, note that representation of
$ A^{(k)}_{l} $ in the form~(\ref{nnn}) is not unique. Indeed,
the expansion of
$ (a-b)^{n_{1}} $ in series in $(1-a)$
$ (1-b) $ will lead to the redefinition of coefficients
$ C^{k}_{n_{1},n_{2},n_{3}} $ and to the change of the
limits of the summation. We use this freedom to represent
$ A^{(k)}_{l} $ in the form in which all calculations for the
case  $ s\leq r $ can be done in the same manner as  those
for $ r\leq s $. Let us consider the eq.~(\ref{defA}). It is
obvious that due to the translation invariance  of function
$\PsiN{x}{y}$
it does not change by the shift
of the variables
$ {\bar y}+{\bar b}-{\bar a}\to {\bar y} $
and
$ {\bar x}\to {\bar x}+{\bar a}-{\bar b} $. Then carrying out
successively all  operations as in the Sec.{\ref{sproof}}, one
arrive at the formulae~(\ref{nnn}), the summation being
carried out in the following range:
$$
n_{1}=m_{1}+m_{2}; \ \ \ \ n_{2}=r+n-k+m_{2}-m_{3}; \ \ \
n_{3}=k+r+m_{1}+m_{3};
$$
\begin{equation}
 0\leq m_{1}\leq \min{[s,r]};\ \ 0\leq m_{2}\leq r;\ \ \
0\leq m_{3}\leq s-m_{1}.
\end{equation}
The calculation of the coefficients
$ {\tilde A}^{k}_{n_{1},n_{2},n_{3}} $ in this case simply repeats
the one given above.

Thus, for
$ 1\leq k\leq m-1 $ one has the required inequality:
$$
|A^{(k)}_{l}|\leq const A^{(0)}_{l}/l^{2}.
$$
 \end{document}